# Study on Cosmic Ray Background Rejection with a 30 m Stand-Alone IACT using Non-parametric Multivariate Methods in a sub-100 GeV Energy Range

Konopelko, A.
*Department of Physics, Purdue University, 525 Northwestern Avenue,
West Lafayette, IN  47907-2036, USA*
Chilingarian, A., Reimers, A.
*Cosmic Ray Division, Alikhanian Physics Institute, Armenia*

**Abstract**

During the last decade ground-based very high-energy γ-ray astronomy achieved a remarkable advancement in the development of the observational technique for the registration and study of γ-ray emission above 100 GeV. It is widely believed that the next step in its future development will be the construction of telescopes of substantially larger size than the currently used 10 m class telescopes. This can drastically improve the sensitivity of the ground-based detectors for γ rays of energy from 10 to 100 GeV. Based on Monte Carlo simulations of the response of a single stand-alone 30 m imaging atmospheric Cherenkov telescope (IACT) the maximal rejection power against background cosmic ray showers for low energy γ-rays was investigated in great detail. An advanced Bayesian multivariate analysis has been applied to the simulated Cherenkov light images of the γ-ray- and proton-induced air showers. The results obtained here quantitatively testify that the separation between the signal and background images degrades substantially at low energies, and consequently the maximum overall quality factor can only be about 3.1 for γ rays in the 10-30 GeV energy range. Various selection criteria as well as optimal combinations of the standard image parameters utilized for effective image separation have been also evaluated.

**I. Introduction**

Development of the instrumentation in the field of very high-energy (VHE) γ-ray astronomy is nowadays primarily motivated by the physics goals that the astrophysical community seeks to attain (Weekes 2004). Among these goals are *(i)* the observation of supernova remnants (SNR), which are the conjectural sources of the VHE γ rays; *(ii)* the continuous study of the physics of jets in active galactic nuclei (AGN); *(iii)* investigation of the morphology and spectra of pulsar wind nebulae (PWN); *(iv)* a wider search for sources of pulsed γ-ray emission in the VHE range to name a few. Such a variety of physics topics are hard to address with a single ground-based instrument. In fact the physical diversity of the γ-ray emission mechanisms requires a similar diversity of the observational approaches and instrumentation for different energy ranges. For instance further observations of AGN and Pulsars necessitate the reduction of instrumental energy threshold down to at least 10-20 GeV, whereas for detection of a SNR a noticeable upgrade of the telescope sensitivity above 100 GeV is more favourable. Ultimately the design of a major ground-based Cherenkov facility for future dedicated γ-ray observations has to conform to many requirements in order to allow an efficient observational strategy given the expected γ-ray fluxes from sources of an entirely different nature.
The High Energy Stereoscopic System (H.E.S.S), which is a system of four 12 m imaging atmospheric Cherenkov telescopes, has been operating for three years in the Khomas Highland of Namibia, close to Windhoek, at 1800 m above sea level (Hofmann 2005). This next-generation instrument for ground-based γ-ray astronomy has an energy threshold of about 100 GeV and a sensitivity of about 1% of the Crab Nebula flux. Such sensitivity is achieved due to good angular resolution (0.1°), good energy resolution (15%), and a stringent rejection of the cosmic ray background using the stereoscopic approach. Similar stereoscopic arrays are currently under construction and final testing in both Arizona, and Woomera, Australia. Two 17 m telescopes are being built by the MAGIC collaboration on the Canary Island of La Palma. One of these has been taking data since fall 2004.

The outstanding physics results obtained with H.E.S.S. and MAGIC in the first few years of their operation are a strong motivation for the further development of the imaging atmospheric Cherenkov technique and is basically driven by a further reduction of the energy threshold for future γ-ray observations. Here we are presenting results for such a detector, a 30 m stand-alone imaging atmospheric Cherenkov telescope (IACT) that may potentially achieve an energy threshold as low as 10 GeV and is representative of a prototype for future low energy telescope arrays (see Konopelko 2005). The performance of such a telescope is basically determined by its efficiency at cosmic ray background rejection in the sub-100 GeV energy range. This important issue will be addressed in this paper using detailed Monte Carlo simulations and advanced statistical analysis methods.

**Table 1.** Basic parameters of the simulation setup.

| | |
|---|---|
| Altitude: | 1.8 km above sea level |
| Atmosphere: | Tropical |
| Reflector size: | 30 m |
| Reflector design: | parabolic (F/D=1.25) |
| Number of camera pixels: | 1951 |
| Pixel size: | $0.07^\circ$ |
| Photon-to-photoelectron efficiency: | 0.1 |
| Trigger: | Signal in each of 3 PMs exceeds 6 ph.-e. |
| 'Boundary'/'Picture' thresholds: | 3/5 ph.-e. |

## 2. Simulations

The atmospheric showers induced by the γ rays and protons have been simulated using the numerical code described in Konopelko et al. (2000). The primary energy of simulated showers was uniformly randomized within each of three energy bins, which were chosen to cover the energy range starting from 1 GeV and extending up to 1 TeV. The events were weighted according to a power-law primary spectrum and the reconstructed shower energy. The maximum impact distance of the shower axis with respect to the centre of the 30 m Cherenkov telescope was 300 m. All showers were simulated at the zenith with a random sampling over azimuth. This reduces any systematic bias in the distributions of the image parameters due to the geomagnetic effect, although it noticeably enhances fluctuations in individual showers. The basic parameters of the simulation setup are summarized in Table 1. The detailed simulation procedure of the camera response accounts for all efficiencies involved in the process of the Cherenkov light propagation, which starts from the photon emission in a shower and ends with the digitization of the camera photomultiplier tube(PMT) signal. This includes the atmospheric absorption, the mirror reflectivity, the photon-to-photoelectron conversion inside the PMT *etc*. The overall efficiency of the photon-to-photoelectron conversion is ~0.1. The standard 'picture' and 'boundary' technique with thresholds of 5 and 3 photoelectrons (ph.-e.), respectively, was applied for image cleaning. The procedure accepts for computation of the second-moment image parameters all camera PMT signals that exceed the 'picture' threshold, and only those PMT signals that exceed the 'boundary' threshold and are adjacent to any of the 'picture' pixels. The simulated images have been parameterized using the standard measures of their angular extension and orientation in the telescope focal plane. Further details on the simulation procedure can be found in Konopelko et al. (1999; 2005).

The basic parameters of the simulation setup have been chosen to meet the major technical requirements for effective imaging of the atmospheric showers. The parabolic optical reflector yields a point-spread function of sufficiently narrow width (~$0.06^\circ$) in the range of the light incidence angles of $1.75^\circ$. This constrains the choice of the minimum angular size of a PMT and the total number of PMT in the camera, which ultimately determines the camera field of view. Note that any increase of the PMT angular size will substantially degrade the image parameterization of the low energy γ rays. At the same time further reduction of the PMT angular size or an increase of the camera field of view will not be beneficial due to significant optical smearing, which is a major limiting factor. All other parameters of the simulation setup such as the atmosphere, the geomagnetic field strength, and the observational height have been chosen to match the environmental conditions of the H.E.S.S. II project (Hofmann 2005), which is a large 28 m IACT under construction in Namibia.

The proton-induced air showers simulated here started at 1 GeV. This energy accounts for all the secondary muons that could trigger the telescope. Here we did not distinguish between the images generated by a single muon or a low energy proton shower that are very similar in shape. That is why, in the sub-100 GeV energy range, the standard anti-muon cut (Length/Size) does not work effectively against muon events but rather increases the energy threshold for γ-ray showers.

It is worth noting that the images of simulated γ-ray- and cosmic-ray-induced atmospheric showers at TeV energies were formerly crosschecked versus the images recorded with the HEGRA (Konopelko et al. 1999) and H.E.S.S. (Konopelko et al. 2003) experiments.

The performance of a single 30 m IACT was discussed in Konopelko (2005) in great detail. Thus the expected raw event detection rate for such a telescope is expected to be about 1.7 kHz. Note that such a high rate can still be maintained by conventional data acquisition systems. The cosmic electrons contribute substantially to this high rate, but the rate is still dominated by the cosmic ray protons and nuclei. Even after applying the standard analysis cuts the remaining proton rate at energies above 20 GeV exceeds the electron rate by a factor of two.

### 3. Bayesian Paradigm

The background rejection strategies can be divided into two major categories:

1. *A-priori* strategies derived from the simulations of both the γ-ray- and cosmic-ray-induced atmospheric showers. For each simulated shower the Cherenkov light image can be generated and parameterized. By varying the energy and impact distance of simulated showers and taking into account all possible distortions in the hardware response, including a trigger decision, image cleaning *etc* one can obtain the so-called training samples for both the γ-ray and cosmic ray primaries. It is in fact very difficult to parameterize the multivariate distribution function for a number of image parameters; therefore we deal with simulation results as they are, i.e. with the samples of the simulated images for the proton- and γ-ray-initiated showers. To represent such a sample of the simulated data we will use, instead of the underlying multivariate distribution function, the special methods of non-parametric statistics.

2. *A-posteriori* strategies based on the experimental data: the so-called on-source sample of events, which were recorded when the telescope was tracking a putative γ-ray source, and the off-source sample, recorded when the telescope was pointed at the same celestial coordinates, but delayed by 28 min after (or before) the source passage. Using these two signal and background samples it becomes possible to pose the problem of searching *a signal domain*: a volume limited by a multi-dimensional non-linear surface, which includes a majority of the signal events and which is substantially enhancing the signal events content and consequently significantly enlarging the so-called *signal-to-noise ratio*. Further details on *a-posteriori* strategy in the γ-ray signal evaluation can be found, for instance, in Chilingarian & Cawley (1991) or Chilingarian (1993).

A first attempt to develop a statistical theory of cosmic-ray background rejection in the framework of the Bayesian approach for the analysis of VHE γ-ray data was undertaken by Aharonian et al (1990; 1991). This statistical theory includes:

- Selection of the optimal subset of parameters for discrimination purposes;
- Introducing the Bayesian decision rules;
- Introducing the P-values of the statistical tests that indicate the overlap between the parameter distributions of two different event classes;
- Correlation analysis revealing the best pairs to be used in the discrimination process;
- Estimation of the Bayes risk (probability of misclassification) as a measure of the overlap of the multivariate distributions;
- Adaptive models of the Parzen and K-Nearest-Neighbor non-parametric density estimation, for a detailed discussion of these models see Parzen (1962) and Tapia (1978).

Generally, to prove the existence of a γ-ray source one calculates the excess of events coming from the direction of a possible source, $N_{on} - N_{off}$. Here $N_{on}$ is a number of events in the on-source sample, which has to be compared to the control event sample. This control sample must guarantee that pure background events have been recorded, $N_{off}$. The expected γ-ray fluxes are often very weak and the signal to background ratio might frequently be very small, less than 0.01. In such a case one should always answer the following generic question: is the detected abundance a real signal or only a background fluctuation? The measure of statistical significance commonly used in VHE γ-ray astronomy is the so-called signal-to-noise ratio, σ (e.g. Zhang et al. 1990):

$$\sigma = \frac{N_{on} - N_{off}}{\sqrt{N_{on} + N_{off}}} \qquad (1)$$

The larger the signal-to-noise ratio (σ) the smaller the probability that the detected excess is due to a background fluctuation. Development of new detector hardware and new data handling methods aim to enlarge the value of σ. After selecting the γ like events from the raw data, in both ON and OFF data samples, the criterion takes the form:

$$\sigma^* = \frac{N^*_{on} - N^*_{off}}{\sqrt{N^*_{on} + N^*_{off}}}, \qquad (2)$$

where $N_{on}^*$, $N_{off}^*$ are the numbers of the ON and OFF events surviving image selection cuts.

Using the actual values of the image parameters measured for each individual event one has to decide whether this event was initiated by a γ ray or cosmic ray. This statistical decision problem in the Bayesian approach can be described in terms of the following probability measures, defined in metric spaces. Let us introduce the set of possible *states of nature* $\mathbf{A} \ldots (\gamma, h)$, e.g. the γ rays ($\gamma$) and cosmic ray hadrons (*h*). The set of all possible statistical decisions is $\tilde{\mathbf{A}} \ldots (\tilde{\gamma}, \tilde{h})$, where the tilde sign denotes the statistical decisions for any examined event, which may belong to one of the signal or background samples. Both decision sets contain the same two elements, but they are not identical: in the first case we deal with *a-priori* given categories, while the second set reflects the results of applying any specific statistical evaluation procedure, constructed for the classification of the experimentally measured events into two given classes.

By multiplication of these two sets we define the so-called loss measure, $c_{A\tilde{A}}$, which indicates the possible consequences of any applied statistical decision. For the problem of background rejection in VHE γ-ray astronomy it is logical to define zero losses for correct classification:

$$c_{\gamma\tilde{\gamma}} = c_{h\tilde{h}} = 0. \qquad (3)$$

If we misclassify a signal event, we decrease the acceptance efficiency for $\gamma$-ray events. At the same time if we erroneously attribute some cosmic ray event to a $\gamma$-ray event, we increase the background contamination. As we initially expect to observe a significant excess of background events over signal events, we are interested in very strong background suppression. Therefore it is reasonable to introduce a non-symmetric loss function for this case, for example:

$$c_{\gamma\tilde{h}} = 0.01, \quad c_{h\tilde{\gamma}} = 0.99. \qquad (4)$$

The dimension of the event entry space, $\mathcal{V}$ (measurements, features *etc*), is defined in our case by a number of measured image parameters. For example, one could measure the number of camera pixels with non-zero signal.

A prior measure $P_A \ldots (P_\gamma, P_h)$ is the assumed proportion of the γ rays and cosmic rays in the raw data flow. The conditional densities or Likelihood functions of image parameters $\mathbf{v} \subset \mathcal{V}$ are denoted as:

$$\hat{p}(\mathbf{v}/\gamma), \hat{p}(\mathbf{v}/h). \qquad (5)$$

These probability density functions can be estimated using the training samples and are in fact the main elements of the decision rule. Multivariate probability density estimation is a fundamental problem in data analysis, pattern recognition, and even artificial intelligence. Naturally, we find the estimation of the conditional density using simulations is a key problem in VHE γ-ray astronomy as well.

## 4. Bayesian Decision Rules

An optimal decision rule should minimize the mean losses, averaged over all possible statistical decisions. For the special selection of the loss function Eqns. (3, 4) the correct statistical decisions will not introduce any losses, therefore we have to select between two possibilities: to erroneously discard a γ-ray image, or to erroneously accept cosmic ray h-images as signal. The non-parametric Bayesian decision rule η depends on the conditional densities, loss functions, and on some prior measures, and has a generic form:

$$\tilde{A} = \eta(\mathbf{v}, A, \tilde{P}) = \arg\{\min_i \{c_i \hat{p}(A_i / \mathbf{v})\}, \ i = \gamma, h\}, \quad (6)$$

where $c_i$ is the loss connected to the decision $\tilde{A}$. $\hat{p}(A_i / \mathbf{v})$ is the non-parametric estimate of *a posteriori* density, which is connected to the conditional density by the Bayes theorem:

$$\tilde{p}(A_i / \mathbf{v}) = \frac{P_i \hat{p}(\mathbf{v} / A_i)}{\hat{p}(\mathbf{v})}, \quad (7)$$

where $\hat{p}(\mathbf{v}) = \hat{p}(\mathbf{v} / \gamma) + \hat{p}(\mathbf{v} / h)$.

Finally, substituting the *a posteriori* densities with the conditional[1] ones we get the Bayesian decision rule in the form:

$$\tilde{A} = \arg\{\min_i \{c_i P_i \hat{p}(\mathbf{v} / A_i)\}, \ i = \gamma, h\} \quad (8)$$

As one can easily see from Eqn. (8) the Bayesian statistical decision depends on the product of $c_i P_i$. Therefore we can not separate the influence of loss measure and prior measure on the decision rule. We will treat the multiplication $c_i P_i$ as a unique term and ascribe it as *a priory loss*. To investigate the influence of chosen values of a priory losses the event type evaluation procedure has been performed simultaneously using various variants of the a priory losses (see below). Examining the so called *influence curves*, obtained for different losses, one can select the preferable regime of the decision rule. For instance it is easy to control the ratio of the background suppression factor to the signal event acceptance.

## 5. Non-parametric Probability Density Estimators

To estimate conditional densities we used here the Parzen method of the probability density estimation (Devroye et al., 1985; Parzen, 1962) with an automatic choice of the kernel width (Chilingarian & Galfayan, 1984). Several estimates of the probability density, which correspond to a number of Parzen kernel widths, were calculated simultaneously. Afterwards the sequence of all derived estimates was ordered according to the magnitude of the signal-to-noise ratio. The median entry of this sequence was chosen as a final estimate. Such estimator of the probability density function (L-estimator) has apparent stabilizing properties for the final estimate by reason that the best estimate is chosen among a number of calculated ones (Efron, 1981).

The Parzen kernel-type probability density is defined as:

$$\hat{p}(\mathbf{v} / A_i) = \frac{|\Sigma_i|^{-0.5}}{(2\pi)^{d/2} s^d} \sum_{j=1}^{M_i} e^{-r_j^2 / 2s^2} \omega_j, \ i = 1, ..., L, \ \sum_{j=1}^{M_i} \omega_j = 1, \quad (9)$$

where $d$ is the dimension of the multi-parameter space, $M_i$ is the number of events in the $i-th$ training sample, $w_j$ are the event weights (e.g. the energy spectrum weights); $s$ is the kernel width (this the only free parameter which controls the smoothness of the estimate), $r_j$ is the distance from the experimentally measured event $\mathbf{v}$ to the $j-th$ event of the training sample, $\mathbf{u_j}$, in the multi-parameter space using the Mahalanobis metric (Mahalonobis, 1936):

$$r_j^2 = (\mathbf{v} - \mathbf{u_j})^T \sum_i^{-1} (\mathbf{v} - \mathbf{u_j}), \qquad (10)$$

where $\sum_i^{-1}$ is the sampling covariance matrix of the event class (γ rays, cosmic rays) to which $\mathbf{u}_j$ belongs.

---

[1] The conditional density *f(x/A)* is the density of a variable *x* given any specific condition *A*, e.g. the type of the primary particle is a hadron.

## 6. Bayes Error Estimation

The most natural measure of both the selection of a best subset of the features and the performance of the event type evaluation is the classification error probability. It depends in turn on both the degree of overlap of a few alternative multivariate distributions and the quality of the decision rule applied. It is worth noting that the Bayes decision rule provides the minimal classification errors as compared to any other decision rule strategy. Bayes errors can be calculated as follows:

$$R^B = E\{\theta[\eta(\mathbf{v}, \mathbf{A}, \mathbf{P})]\} = \int \theta p(\mathbf{v}) d\mathbf{v} \qquad (11)$$

where

$$\theta[\eta(\mathbf{v}, \mathbf{A}, \mathbf{P})] = \begin{cases} 0, \text{ for correct classification} \\ 1, \text{ otherwise} \end{cases} \qquad (12)$$

and $\eta(\mathbf{v}, \mathcal{A}, \mathcal{P})$ is the decision rule defined by Eqn. (8).

The mathematical average is calculated for the whole $d$- dimensional feature space $\mathcal{V}$. In other words the Bayes error is a measure of the overlap of alternative distributions of different event classes in the feature space $\mathcal{V}$, e.g. it gives a relative contribution of all *incorrectly* classified events. Since we do not know to which event class each particular individual event recorded in the experiment belongs, we can obtain an estimate of $R^B$ exclusively using the training samples $\mathbf{u}_j$ :

$$\widehat{R}^B = E\left\{ \frac{1}{M_{TS}} \sum_{i=1}^{M_{TS}} \theta\left[\eta\left(\mathbf{u_i}, \mathbf{A}, \tilde{\mathbf{P}}\right)\right] \right\}, \qquad (13)$$

i.e. we classify the simulated events $\{\mathbf{u}_i\}, i = 1, M_{TS}$ and check the correctness of the classification. An average error is calculated over all possible samples of size $M_{TS}$. Many independent investigations have shown (e.g. Toussaint, 1974) that this estimate is systematically biased and hence, the so-called *one-leave-out-for-a-time* estimate is preferable

$$\hat{R}^e = \frac{1}{M_{TS}} \sum_{i=1}^{M_{TS}} \theta\{\eta(\mathbf{u_i}, \mathbf{A}, \tilde{\mathbf{P}}_{(i)})\} \qquad (14)$$

where $\left(\mathbf{A}, \tilde{\mathbf{P}}_{(i)}\right)$ is a training sample without the $i$-th element, which is classified first and then *returned* back into the sample. This estimate is unbiased and essentially has a smaller mean squared deviation compared to other estimators (Snappin et al., 1984). The advantage of $\hat{R}^e$ is especially notable when the feature space is of high dimension. Note that we can estimate the erroneous classification probability by classifying various Training Sample classes. In this way we can estimate the expected γ-ray event acceptance efficiency and the cosmic ray contamination.

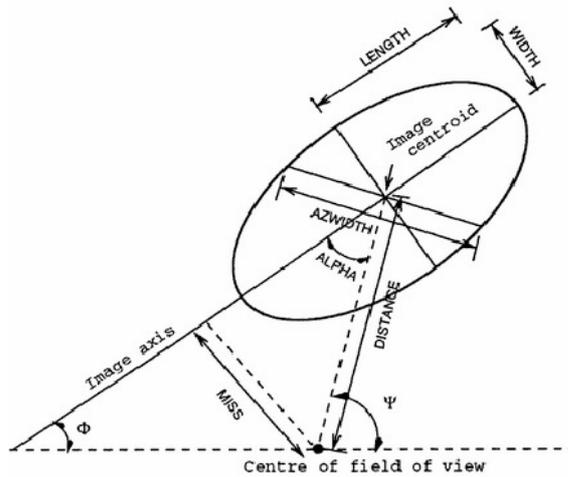

Figure 1. The second-moment parameters of the Cherenkov light image.

## 7. Estimates of the Background Rejection Rates

After selecting the best single discriminate out of the image parameters (see Figure 1) and the best pairs of discriminates using the technique first developed by Aharonian et al. (1991), which is implemented in the applied statistical decision package ANI (Chilingarian, 1989; Chilingarian, 1998), the same parameters used for the image analysis of data taken with the 10 m Whipple Collaboration telescope were proven to be the best (Aharonian, 1991). They are the image shape parameters Width, Length, and the combined parameter of image shape and orientation, AzWidth. At the same time, as shown in Figure 2, the single shape parameter Width cannot provide any significant discrimination for both the low and high-energy γ rays considered here.

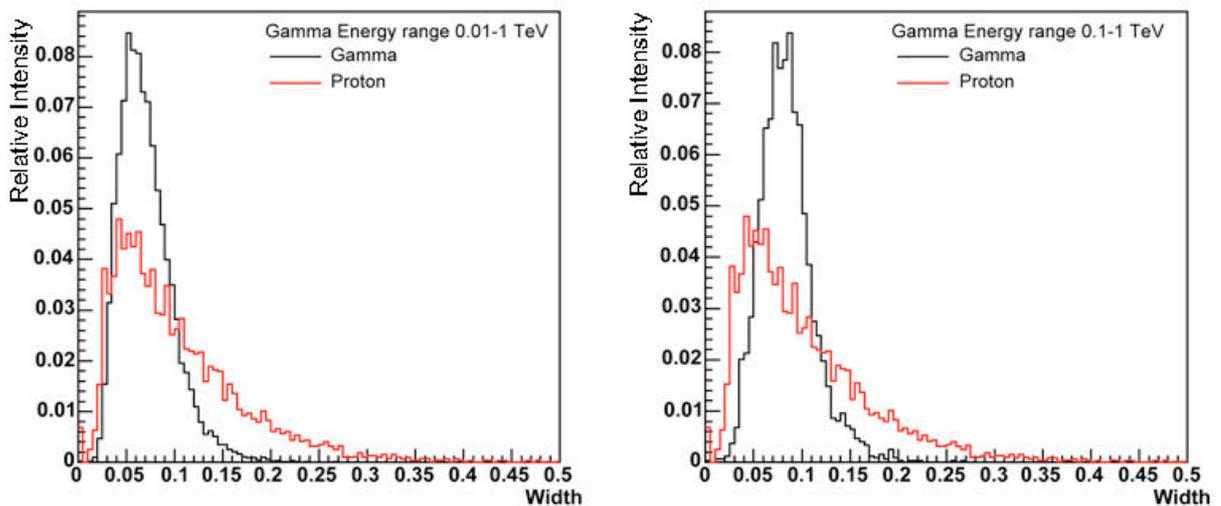

Figure 2. Illustration of the discrimination power between γ rays and protons using a single image parameter, Width, in two energy intervals.

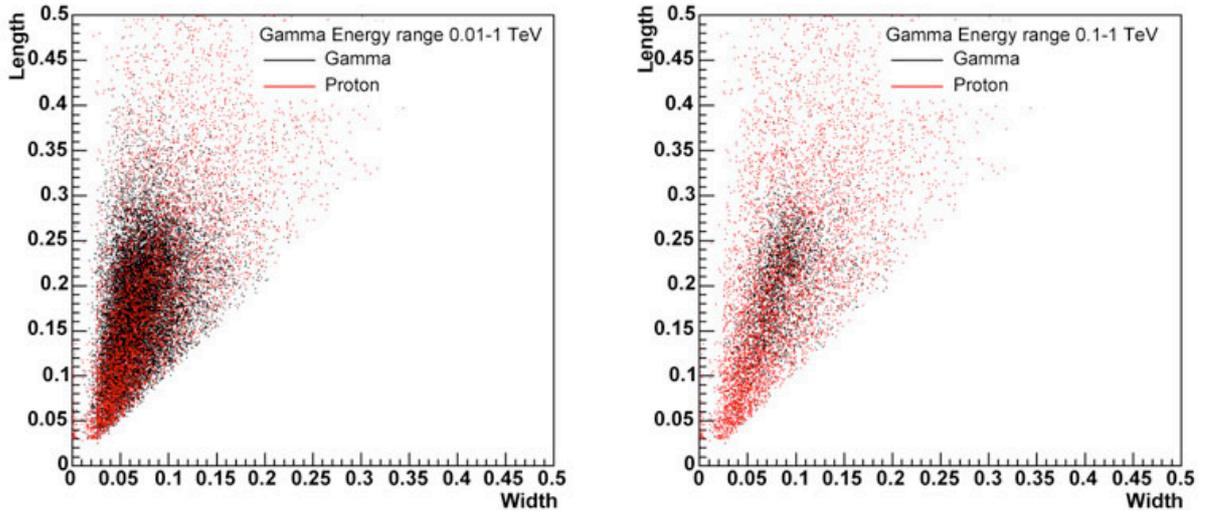

Figure 3. Illustration of the discrimination power between γ rays and protons using simultaneously two image parameters, Width and Length, in two energy intervals.

Adding the second shape parameter Length significantly improves the situation. One can see in Figure 3 that for the images of high-energy events we can outline a two-dimensional domain where most of the γ like events are included. Apparently the low energy interval contains much more discrepant and diffuse images and is widely spread in the parameter space. As a result the discrimination is much worse for those events as compared to the high-energy interval.

An additional orientation parameter Alpha taken along with the shape parameters Width and Length substantially improves the situation and we can see compact γ-ray domains for both high and low energy events (see Figure 4). Again the γ-ray domain for low energy events is much larger. Nevertheless, the observed concentration of the γ-ray events might allow a treatment of the background rejection problem even for the low energy events.

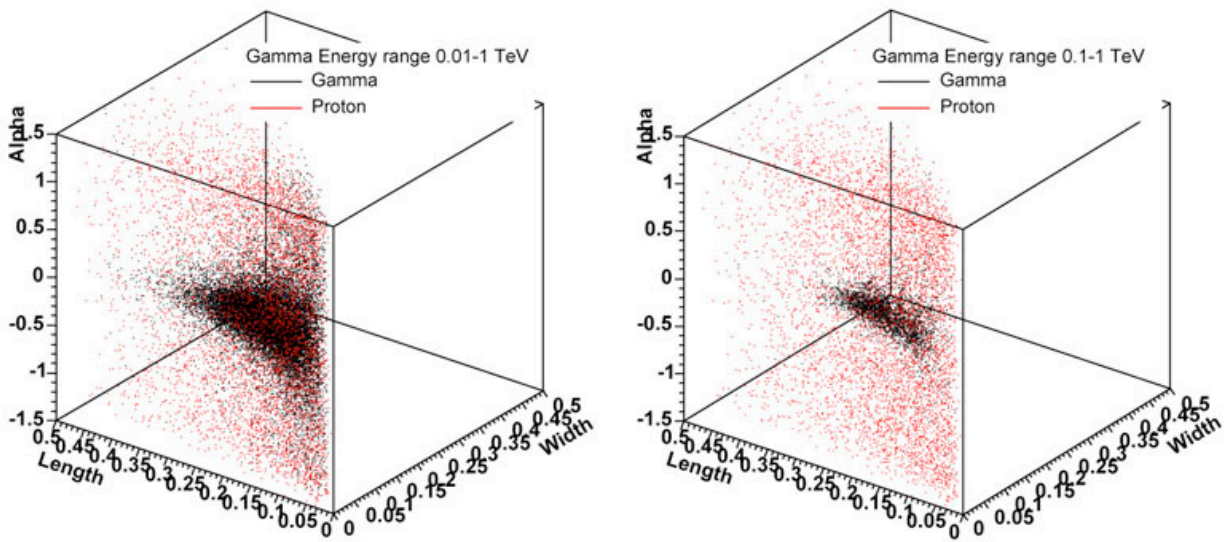

Figure 4. The γ-ray domain in the three-dimensional space of the image parameters.

In Figure 5 the best set of image parameters is shown. After applying a one-dimensional analysis, correlation analysis, and the Bhattacharia distance minimization technique described by Aharonian et al. (1991) we performed multiple calculations of the Bayes Risk given in Eqn. (13). Here we applied the

Bayes decision rule Eqn. (8) using the numerical approximation of the probability density function given in Eqn. (9).

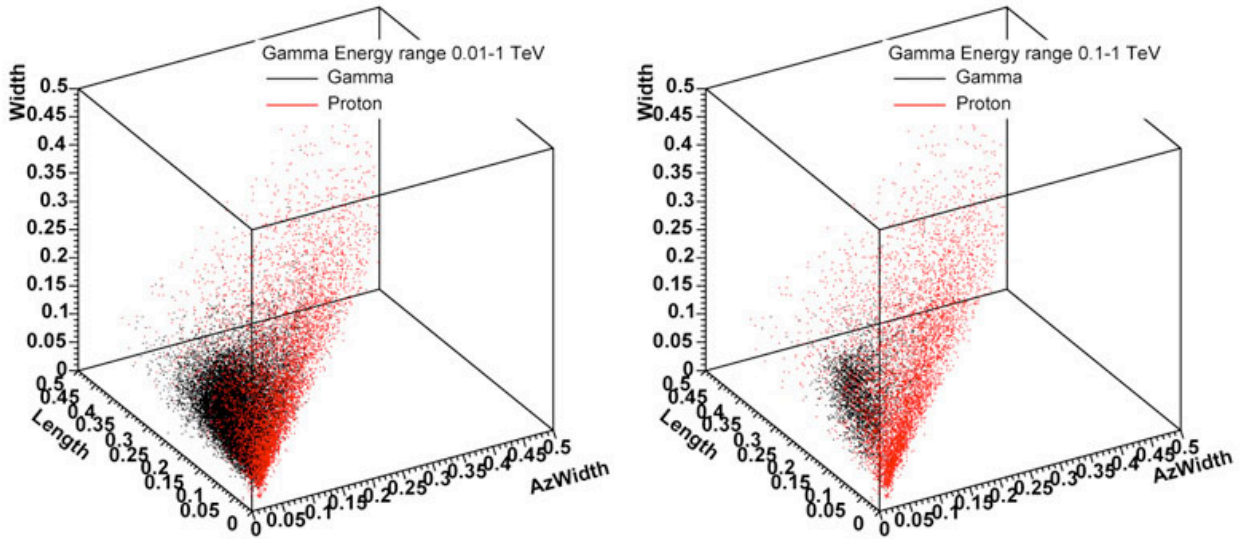

Figure 5. The γ-ray domain in the three-dimensional space of the image parameters. This domain provides a minimal Bayes error.

The results of calculations using different loss functions and two different energy intervals are summarized in Figure 6. In the present analysis we used the overlapping energy intervals but we have utilized the reconstructed energy for each individual event. For a power-law spectrum most of the triggered events occur right above the energy threshold, e.g. 10 or 100 GeV, so such analysis clearly illustrates how rejection power depends on shower energy. Figure 7 shows the so-called *influence curves*. By varying the $c_i$ parameter in Eq. (6) one can obtain the classification results for a number of different loss functions. It is the way to obtain the influence curve. The influence curve displays all possible combinations of the γ-ray acceptance efficiency and the corresponding cosmic ray background contamination. For instance one can obtain a very high efficiency at the cost of a rather large background contamination or vice-versa. For the high-energy interval it is possible to achieve a cosmic ray background rejection much less than one percent while keeping the acceptance of γ rays at about 50%. The situation dramatically worsens at low energy. However the discrimination is still possible even for the lowest energy interval of 10-30 GeV, where we can obtain a so-called Q-factor ($Q = \varepsilon_\gamma / \sqrt{\varepsilon_h}$, where $\varepsilon_\gamma$ is the γ-ray acceptance efficiency and $\varepsilon_h$ is the corresponding cosmic ray hadron contamination) of about 3.1. For higher energy ranges, 30-50 GeV and 50-100 GeV, the Q-factors are 3.8 and 4.2, respectively (see Figure 7). It is worth noting that the image parameter AzWidth is directly related to the image orientation and the angular resolution of the telescope, which is about 0.3° at low energies. Such modest angular resolution is a result of the large fluctuations in development of the low energy shower, the geomagnetic field deflection of shower electrons, and the rather low photoelectron content of these images. Note that currently achieved telescope pointing accuracy is less than 1 arc min, which is negligible compared with the actual angular resolution of the low energy γ rays.

## Summary

The construction of a 30 m telescope in Namibia is currently ongoing. Such an instrument is strongly supported by various physics motivations for studying γ-rays around and above 20 GeV. The performance of this instrument greatly relies on the ability to extract γ-ray signal out of the dominant cosmic ray background. Any possible further advancement in the analysis that may improve the performance of future γ-ray observatories is very important. Here we applied a multivariate analysis to the simulated data for a single stand-alone 30 m imaging Cherenkov telescope. Despite the fact that we have used a complete Monte Carlo simulation of the telescope response a few minor effects that might distort the performance of the telescope were not taken into account, such as detailed timing of reflected photons etc. However the major results on the classification efficiency for a 30 m telescope of a specific hardware design should not largely deviate from the results obtained here.

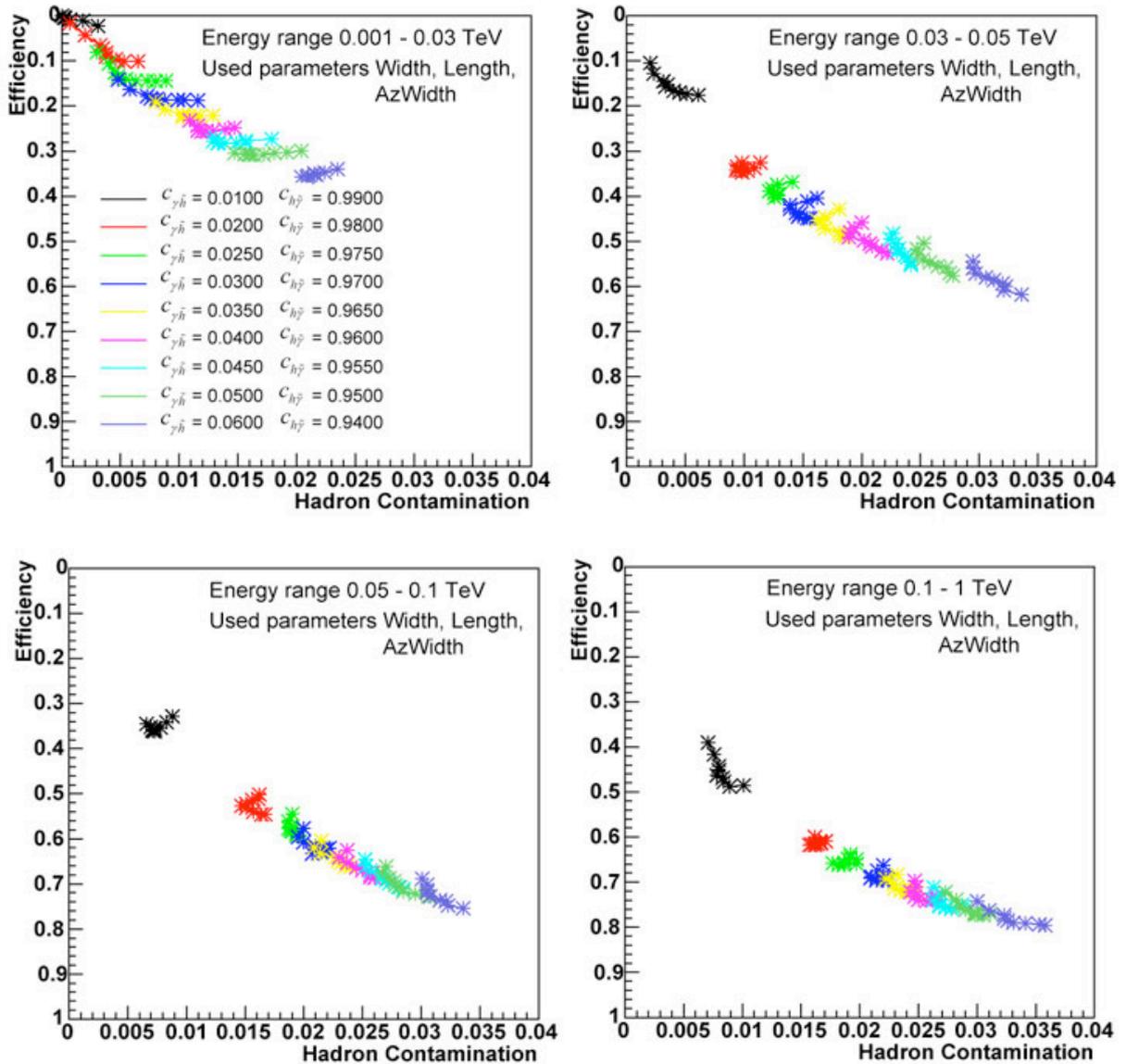

Figure 6. Comparison of the Bayesian risk estimates for four different energy intervals

A single imaging Cherenkov telescope does not allow very accurate measurement of the arrival direction of individual showers. A number of currently existing advanced methods of shower reconstruction for a single telescope become totally ineffective at low energies, such as 10 GeV. This fact can be explained by large fluctuations in the shape and orientation of the low energy images. Motivated by that, we applied here a set of image parameters that utilizes the difference in correlations between two basic parameters, Width and AzWidth, for the γ-ray and cosmic ray showers. One can directly apply the recommended combination of the image parameters to the forthcoming experimental data.

The results shown in Figure 6 demonstrate that the discrimination power against a cosmic ray background substantially worsens at low energies. The maximum quality factor obtained for the γ-ray events in the energy range of 10-30 GeV is about 3.1. This value is significantly lower than the value achieved for the energy range above 100 GeV, which is about 7. Given a very steep predicted spectrum of γ-ray emission in the sub-100 GeV energy range for various sources (e.g. Pulsars, distant AGN *etc*) a relatively modest rejection power could be compensated for by the enhanced statistics of the γ rays. The situation could be substantially improved for a system of 3 to 5 identical 30 m telescopes operating in stereoscopic mode.

Discussion of an advanced multi-telescope analysis for observations of low energy γ-rays with a system of a few 30 m IACTs will be the subject of a forthcoming paper.

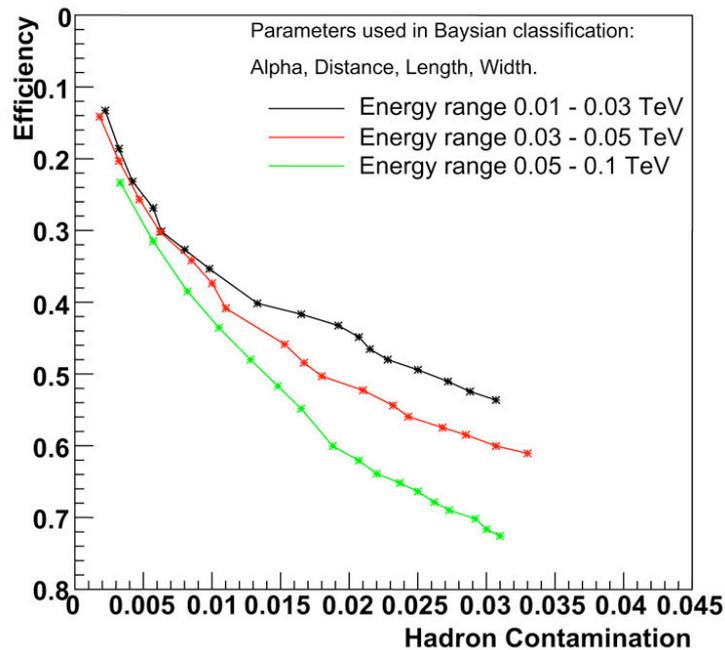

Figure 7. The influence curves for three different energy ranges of the simulated γ-rays.

## Acknowledgements

AK and AC are very thankful to Nikolai Pavel* for support of this project. AC would like also to thank Nikolai Pavel and all members of his group for hospitality during his stay at Institute of Physics of Humboldt-University of Berlin. AK would like to thank John P. Finley for discussions on a subject of this paper.

## References

1. Aharonian F. A., Chilingaryan A. A., Konopelko A. K., Plyasheshnikov A. V. On the possibility of an improvement of background hadronic showers discrimination against γ-ray coming from a discrete source by a multidimensional Cherenkov light analysis, *Proc. 21 ICRC*, vol. 4, p. 246, Adelaide, 1990.

2. Aharonian F.A., Chilingarian A.A., Konopelko A. K., Plyasheshnikov A. V. A multidimensional analysis of the Cherenkov images of air showers induced by very high-energy γ-rays and protons, *NIM* A302, 522, 1991; *YerPhI preprint* 1171 (48)-79, 1989.

3. Chilingarian A. A., Statistical decisions under nonparametric *a priory* information, *Computer Physics Comunication*, 54, 381-390, 1989

4. Chilingarian, A.A., Cawley, M.F. Application of multivariate analysis to atmospheric Cherenkov imaging data from the Crab nebula. *Proc. 22 ICRC*, 1, 460-463, Dublin, 1991

5. Chilingarian A. A., Galfayan S. Kh., Calculation of Bayes risk by KNN method, Stat. *Problems of Control*, 66, 66, Vilnius, 1984

6. Chilingarian A. A. On the methods of the enhancement of the reliability of the signal detection with Cherenkov Atmospheric techniques, *Izv. AN USSR*, Phys., vol. 57, p. 186, 1993

7. Chilingarian A. A., Analysis and Nonparametric Inference in High Energy Physics and Astroparticle Physics, 1998, program Package ANI, (User's Manual, unpublished) http: //crdlx5.yerphi.am/proj/ani

---

* Deceased

8. Devroye L., Gyorfi L., Nonparametric density estimation. The LI view, John Wiley and Sons, New-York, 1985

9. Efron B., Nonparametric standard errors and confidence intervals, *Canadian J. Statist.*, 9, 139, 1981

10. Fix E., Hodges J. L. Discriminatory analysis. Nonparametric discrimination, Consistency Properties, *Project 21-49-004, Report 4,USAF School of Aviation Medicine*, 1951

11. Fukunaga K., Himmels D., Bayes error estimation using Parzen and KNN procedures, IEEE Trans., PAMI-9, 634, 1987

12. Hofmann, W. *Proc. 29 Int. Cosmic Ray Conference*, Pune, India, Vol. 10, 97-114, 2005

13. Konopelko, A. Performance of the stereoscopic system of the HEGRA imaging air Cerenkov telescopes: Monte Carlo simulations and observations, *Astroparticle Physics*, Volume 10, Issue 4, p. 275, 1999

14. Konopelko, A., Plyasheshnikov, A.V. ALTAI: computational code for the simulations of TeV air showers as observed with the ground-based imaging atmospheric Cherenkov telescopes, *Nucl. Instr. & Meth. in Phys. Res*. Section A, Volume 450, Issue 2-3, p. 419-429, 2000

15. Konopelko, A, et al., for the H.E.S.S. collaboration *Proc. 28th Int. Cosmic Ray Conf.*, Tsukuba, Univ. Academy Press, Tokyo, p. 2903, 2003

16. Konopelko, A. STEREO ARRAY of 30 m imaging atmospheric Čerenkov telescopes: A next-generation detector for ground-based high energy γ-ray astronomy *Astroparticle Physics*, Volume 24, Issue 3, p. 191-207, 2005

17. Lofsgaarden D. O., Quesenberry C. D., A nonparametric estimate of a multi-variate density function*, Ann. Math. Stat.*, 36, 1049, 1966

18. Mahalonobis P. C. On the generalized distance in statistics, *National Inst. of India*, 2, 49, 1936

19. Parzen E. On estimation of a probability density function and mode, *Ann. Math. Stat.*, 33, 1065, 1962

20. Rabiner L. R., Levinson E., Rozenberg A. E., Wilpon J, G. Speaker - independent recognition of isolated words using clustering techniques*, IEEE Trans, on Acoustics, Speech, Signal Processing*, ASSP-27, 336, 1974

21. Rosenblatt M. Remarks on some nonparametric estimates of a density function, *Ann. Math. Stat.*, 27, 832, 1957

22. Snappin S. M., Knoke J. D. Classification error rate estimators evaluated by unconditional mean squared error, *Technometrics*, 26, 371, 1984

23. Tapia R. A., Thompson J. R. Nonparametric probability density estimation, The John Hopkins University Press, Baltimore and London, 1978

24. Toussaint G. T., Bibliography of misclassification, IEEE trans. on Information, p. 472, 1974

25. Zhang, S.N., Ramsden, D. Statistical data analysis for γ-ray astronomy. *Exp. Astronomy* 1, 147, 1990

26. Weekes, T.C. Very High Energy Gamma Ray Astronomy, Inst. of Phys., Series in Astron. & Astrophys., IOP Publishing Ltd, 2003